\documentclass[12pt]{article}                 

\usepackage[latin1]{inputenc}
\usepackage[english]{babel}
\usepackage[T1]{fontenc}
\usepackage{amsmath}
\usepackage{amsfonts}
\usepackage{a4wide}
\usepackage{graphicx}
\usepackage{textcomp}
\usepackage{ae}
\usepackage[colorlinks=true, linkcolor=blue, citecolor=blue, urlcolor=blue]{hyperref}
\usepackage{color}
\usepackage{cite}
\usepackage{titlesec}
\usepackage{fancyhdr}
\usepackage{url}
\usepackage{soul}

\begin{document}
\title{Influence of the Raman depolarisation ratio on far-field radiation patterns in coherent anti-Stokes Raman scattering (CARS) microscopy}
\author{David Gachet\footnote{david.gachet@fresnel.fr}, Nicolas Sandeau\footnote{nicolas.sandeau@fresnel.fr}, Hervé Rigneault\footnote{herve.rigneault@fresnel.fr}}
\date{\small{received June 29, 2006; published September 12, 2006}}
\maketitle

\begin{center}
\emph{Institut Fresnel, Mosaic group, UMR CNRS 6133, Université Paul Cézanne Aix-Marseille III, Domaine Universitaire St-Jérôme F-13397 Marseille cedex 20, France}
\url{http://www.fresnel.fr/mosaic/}

\end{center}

We propose a full-vectorial numerical study of far-field radiation patterns in coherent anti-Stokes Raman scattering (CARS) microscopy. We emphasis the particular role of the Raman depolarisation ratio of the observed medium and show how it modifies the radiation pattern of thin objects.\\
\textbf{Keywords:} CARS, coherent microscopy, CARS electromagnetics modelling.
\section{Introduction}

	  Predicted in 1965 by Maker and Therune \cite{Maker65_pr}, coherent anti-Stokes Raman scattering (CARS) was first used in microscopy as a contrast mechanism in 1982 by Duncan et al. \cite{Duncan82_ol}. It has revealed to be a powerful non-invasive tool to probe the chemical composition of microscopic objects. Since its renewal in 1999 under collinear configuration \cite{Zumbusch99_prl}, it has been considered as valuable to image biological samples. CARS is a third-order nonlinear effect in which two beams at optical pulsations $\omega_p$ and $\omega_s$ mix in the studied medium to give rise to the so-called anti-Stokes radiation at pulsation $\omega_{as}=2\omega_p-\omega_s$. When $\omega_p-\omega_s$ equals a vibrational pulsation of the medium, the anti-Stokes signal is enhanced and vibrational contrast is thus generated. CARS is classically described by the third order nonlinear tensor $\chi^{(3)}$. As a coherent process, CARS efficient generation is very sensitive to the phase-mismatch $\Delta\mathbf{k}=2\mathbf{k_p}-\mathbf{k_s}$. Several geometries have been proposed to relax the phase-matching condition, among them collinear geometry under tight focusing condition \cite{Bjorklund75_jqe} and BOXCARS geometry \cite{Müller00_Jmicro}, the former being the most implemented nowadays. To provide good axial and lateral resolutions, microscope objectives with high numerical apertures (NA) are commonly used \cite{Müller02_jpc,Wang05_biophysj,Potma06_ol}.\\

As a coherent process, CARS generation is very sensitive to both size and shape of imaged objects. Image formation process in CARS microscopy is thus narrowly bound to the far-field CARS radiation pattern of studied samples, so that images are not the simple convolution of the object with the microscope point spread function as in fluorescence microscopy. Moreover, under tight focusing condition, the commonly used paraxial approximation breaks. Hashimoto and al. \cite{Hashimoto01_josa} first derived the coherent transfer function and the optical transfer function of a CARS microscope under scalar assumption. Based on the framework developed by Richards and Wolf to treat the problem of tightly focused beams \cite{Richards59}, Volkmer and al. solved the problem introducing Hertzian dipoles \cite{Volkmer01_prl} and later, Cheng and al. used a Green's function formalism \cite{Cheng02_josa1}. These two studies  took into account (i) the vectorial nature of the pump and Stokes exciting fields and (ii) both the size and the shape of the imaged object. However, assumptions on the exciting fields polarisation near the objective focus were made. In particular, their longitudinal components (along the optical axis) were neglected, neglecting in the same time the longitudinal component of the nonlinear polarisation responsible for CARS radiation.\\

 As we will show in this paper, the relative amplitude of this component is a function of (i) the way the incident beams are focused into the sample and (ii) the Raman depolarisation ratio ($\rho_{R}$) of the probed medium. In other circonstances, this ratio is found to play an important role in elimination of the non-specific signal in polarisation CARS spectroscopy \cite{Akhmanov78_jetp,Oudar79_apl} and microscopy \cite{Cheng01_ol}. The nonlinear polarisation acting as a source for the anti-Stokes CARS generation, introducing its longitudinal component potentially affects the far-field radiation pattern of the studied sample.\\
 
 This paper starts with some CARS basics and expressions of the induced nonlinear polarisation as a function of the Raman depolarisation ratio are derived. In a second part, the computation method used in this paper is briefly described, acompanied by a description of the simulated physical situation. Then, exciting fields and nonlinear polarisations are computed for different focusing conditions and different values for the Raman depolarisation ratio. Finally, the influence of these parameters on far-field CARS radiation patterns, for different classes of objects, are presented.\\ 
 
\section{Raman and CARS background}

CARS is governed by the third order nonlinear tensor $\chi^{(3)}$. It is the superposition of a vibrational resonant term, referred as $\chi^{(3)}_R$ and an electronic nonresonant term denoted $\chi^{(3)}_{NR}$ \cite{Shen84}. At exact vibrational resonance, the former is a purely imaginary number while the latter can be considered as real \cite{Lotem76_pr}. In the case of an isotropic medium, $\chi^{(3)}$ depends on three independent components $\chi^{(3)}_{xxyy}$, $\chi^{(3)}_{xyxy}$ and $\chi^{(3)}_{xyyx}$. Typical CARS experiments involving only two input beams, the pump field is frequency-degenerated. As a consequence, the number of its independent components reduces to two so that \cite{Popov95}
 
\begin{eqnarray}\label{chi_components}
	\chi^{(3)}_{ijkl}=\chi^{(3)}_{xxyy}(\delta_{ij}\delta_{kl}+\delta_{ik}\delta_{jl})+\chi^{(3)}_{xyyx}\delta_{il}\delta_{jk}
\end{eqnarray}

where subscripts \textit{i}, \textit{j}, \textit{k} and \textit{l} refer to cartesian coordinates \textit{x}, \textit{y} or \textit{z}, and $\delta$ refers to the Kronecker delta function.\\

 The link between the two components $\chi^{(3)}_{xxyy}$ and $\chi^{(3)}_{xyyx}$ is not straightforward and it is useful to connect their values with experimental data obtained with spontaneous Raman spectroscopy. It is well known that in spontaneous Raman spectroscopy, a depolarisation ratio \textit{$\rho_{R}$} can be defined. It refers to the faculty of the probed Raman line to depolarise the polarised excitation beam. It is defined by
 
\begin{eqnarray}
	\rho_{R}=\frac{I_s(\bot)}{I_s(\|)}
\end{eqnarray}

where $I_s(\|)$ and $I_s(\bot)$ refer to Stokes intensity respectively polarised parallel and perpendicular to the excitation polarisation. \textit{$\rho_{R}$} is specific to both the probed Raman line and the excitation conditions so that it can be expressed as a function of intrinsic parameters of the Raman line \cite{Yuratich77_mp}

\begin{eqnarray}
	\rho_{R}=\frac{5\gamma_a^2+3\gamma_s^2}{45\alpha^2+4\gamma_s^2}
\end{eqnarray}

where $\alpha$, $\gamma_s$ and $\gamma_a$ respectively refer to the isotropy, and the symmetric and antisymmetric parts of the anisotropy of the usual Raman tensor. By analogy, $\bar{\alpha}$, $\bar{\gamma_s}$ and $\bar{\gamma_a}$ coefficents can be defined for CARS scattering \cite{Yuratich77_mp}.\\

In the same way, a CARS depolarisation ratio $\rho_{CARS}$ is defined by \cite{Yuratich77_mp}

\begin{eqnarray}
	\rho_{CARS}=\frac{\chi^{(3)}_{xyyx}}{\chi^{(3)}_{xxxx}}=\frac{\chi^{(3)}_{xyyx}}{2\chi^{(3)}_{xxyy}+\chi^{(3)}_{xyyx}}.
\end{eqnarray} 

 $\rho_{CARS}$ is simply related to the $\bar{\alpha}$, $\bar{\gamma_s}$ and $\bar{\gamma_a}$ coefficients by the relation
 
\begin{eqnarray}
	\left|\rho_{CARS}\right|^2=\left|\frac{-5\bar{\gamma_a}^2+3\bar{\gamma_s}^2}{45\bar{\alpha}^2+4\bar{\gamma_s}^2}\right|^2.	  
\end{eqnarray}

 In the case when no direct electronic absorption occurs, $\bar{\gamma_a}$ equals zero and both $\bar{\alpha}$ and $\bar{\gamma_s}$ are real. Moreover $\bar{\alpha}$ and $\bar{\gamma_s}$ can be safely identified to their spontaneous counterparts (ie $\bar{\alpha}=\alpha$ and $\bar{\gamma_s}=\gamma_s$) \cite{Otto01_jrs}. Therefore $\rho_{CARS}$ can be assumed to equal $\rho_{R}$. When $\alpha$ equals zero, the Raman line is told to be depolarised and $\rho_{CARS}$ equals $0.75$. In the opposite case ($\gamma_s=0$), the Raman line is totally polarised so that $\rho_{CARS}$ equals $0$. Finally, under the assumption of no electronic absorption from the medium, $\rho_{CARS}$ lies between 0 and 0.75. In the particular case of nonresonant CARS, $\chi^{(3)}_{xxyy}$ equals $\chi^{(3)}_{xyyx}$ in virtue of Kleinman's symmetry \cite{Kleinman62_pr}, and $\rho_{CARS}$ equals $1/3$.\\
 
 Expressing $\chi^{(3)}_{ijkl}$ as a function of $\chi^{(3)}_{xxyy}$ and $\rho_R$, it is straightforward to write
 
\begin{eqnarray}\label{chi_components_bis}	\chi^{(3)}_{ijkl}=\chi^{(3)}_{xxyy}(\delta_{ij}\delta_{kl}+\delta_{ik}\delta_{jl}+\frac{2\rho_R}{1-\rho_R}\delta_{il}\delta_{jk}).
\end{eqnarray}

 The local third-order nonlinear polarisation induced, at the point $\mathbf{r}$, by the pump and the Stokes fields $\mathbf{E_p}$ and $\mathbf{E_s}$ is expressed by

\begin{eqnarray}\label{CARS_polar}						\mathbf{P}^{(3)}(\mathbf{r},-\omega_{as})=\chi^{(3)}(-\omega_{as};\omega_p,\omega_p,-\omega_s)\mathbf{E}_p(\mathbf{r},\omega_p):\mathbf{E}_p(\mathbf{r},\omega_p):\mathbf{E}_s^*(\mathbf{r},-\omega_s)
\end{eqnarray}

where $\omega_p$, $\omega_s$ and $\omega_{as}$ are the respective angular frequencies of the pump, Stokes and anti-Stokes fields, the symbol $^*$ is used for the complex conjugation and the symbol $:$ indicates tensorial product. This nonlinear polarisation is the source of the anti-Stokes field.\\

 Taking into account the pump field frequency-degeneracy and omitting frequency arguments $\omega_p$, $\omega_s$ and $\omega_{as}$, the i-th component (\textit{i}=\textit{x},\textit{y},\textit{z}) $P^{(3)}_i$ of the nonlinear polarisation $\mathbf{P}^{(3)}$ can be expressed as
 
\begin{eqnarray}						
P^{(3)}_i(\mathbf{r})=3\sum_{j,k,l}\chi^{(3)}_{ijkl}E_{p_j}(\mathbf{r})E_{p_k}(\mathbf{r})E_{s_l}^*(\mathbf{r})
\end{eqnarray}

where the subscripts \textit{j}, \textit{k}, \textit{l} run over \textit{x}, \textit{y}, \textit{z}.\\

\section{Computing method and simulated physical situation}

 We have investigated the effects of tightly focused excitation beams on CARS generation with a fully vectorial model. The full description of this model can be found in reference \cite{Gachet06_spie}. For convenience, we briefly sum up its main features. It bases on the framework developped by Richards and Wolf \cite{Richards59} to treat cases when the paraxial approximation breaks. Exciting pump and Stokes beams are assumed to be gaussian and are described as a superposition of plane waves that are focused through a high numerical aperture (NA) microscope objective. The finite size of the back aperture of the objective was also taken into account via a parameter $\beta$. First proposed by Hess et al. \cite{Hess02_biophysj}, it equals the ratio of the back aperture radius $r_0$ to the half width at half maximum (HWHM) $\sigma$ of the gaussian incident beams. The resultant electric fields $\mathbf{E_p}$ and $\mathbf{E_s}$, considered as vectorial, are then computed in the vicinity of the focal plane. They induce dipoles in the active medium (ie the medium emitting CARS radiation), which orientation, phase and strenght are determined by the mean of Eq.(\ref{CARS_polar}). These dipoles act as sources for CARS radiation, which far-field radiation pattern is finally computed.\\
 
\begin{figure}[!ht]
	\centering
		\includegraphics[width=0.6\textwidth]{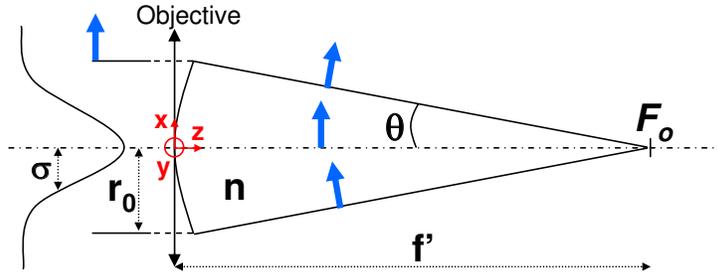}
	\caption{\small{Schematic of the simulated configuration. f': focal distance of the objective; $F_0$: objective's focus; $r_0$: radius of the back aperture of the objective; $\sigma$: incident gaussian beam half-width at half-maximum; n: immersion liquid refractive index ; $\theta$: angle made by extreme rays with the optical axis. The incident beam is linearly polarised (blue arrow) along the x-axis. The parameter $\beta$ is defined by $\beta=r_0/\sigma$.}}
	\label{fig:setup}
\end{figure}

 CARS generation is studied when the whole vectorial components of the electric fields $\mathbf{E_p}$ and $\mathbf{E_s}$ are taken into account in the active medium, which is assumed to be isotropic so that Eq.(\ref{chi_components}) holds. As shown on Figure \ref{fig:setup}, the incident pump and Stokes beam (associated respectively to electric fields $\mathbf{E_p}$ and $\mathbf{E_s}$) are focused in the active medium through a microscope objective (NA=1.2 in water). They are supposed to follow a gaussian spatial distribution, to propagate along the \textit{z}-axis and to be linearly polarised along the \textit{x}-axis (blue arrows on Figure \ref{fig:setup}) so that they are polarised along the \textit{x}- and \textit{z}-axes (the component along the \textit{y}-axis vanishes following \cite{Richards59}) in the vicinity of the focal plane. The higher the angle $\theta$ (see Figure \ref{fig:setup}), and hence the numerical aperture, the stronger the fields components along the \textit{z}-axis. The induced third order nonlinear polarisations along \textit{x}- and \textit{z}-axes equal
 
\begin{eqnarray}	P^{(3)}_x(\mathbf{r})=3\left[\chi^{(3)}_{xxxx}E^2_{p_x}(\mathbf{r})E_{s_x}^*(\mathbf{r})+\chi^{(3)}_{xzzx}E^2_{p_z}(\mathbf{r})E_{s_x}^*(\mathbf{r}) + 2\chi^{(3)}_{xxzz}E_{p_x}(\mathbf{r})E_{p_z}(\mathbf{r})E_{s_z}^*(\mathbf{r})\right]\\
P^{(3)}_z(\mathbf{r})=3\left[\chi^{(3)}_{zzzz}E^2_{p_z}(\mathbf{r})E_{s_z}^*(\mathbf{r})+\chi^{(3)}_{zxxz}E^2_{p_x}(\mathbf{r})E_{s_z}^*(\mathbf{r}) + 2\chi^{(3)}_{zzxx}E_{p_z}(\mathbf{r})E_{p_x}(\mathbf{r})E_{s_x}^*(\mathbf{r})\right]
\end{eqnarray}
 
 The pump and Stokes beams wavelengths are assumed to be respectively $750$ nm and $830$ nm. These values match most of the recent CARS microscopy experiments, where near infrared (NIR) lasers are used \cite{Volkmer05_ap}. Such wavelengths prevent most active media from one and two-photon electronic absorption, so that identification of $\rho_{CARS}$ to $\rho_{R}$ holds and Eq.(\ref{chi_components_bis}) is valid. Therefore, $P^{(3)}_x(\mathbf{r})$ and $P^{(3)}_z(\mathbf{r})$ can be recast under the more convenient form
 
\begin{eqnarray}\label{P_rho_x}	P^{(3)}_x(\mathbf{r},\rho_R)=6\chi^{(3)}_{xxyy}\left\{\frac{1}{1-\rho_R}\left[E^2_{p_x}(\mathbf{r})+\rho_R E^2_{p_z}(\mathbf{r})\right]E_{s_x}^*(\mathbf{r})+E_{p_x}(\mathbf{r})E_{p_z}(\mathbf{r})E_{s_z}^*(\mathbf{r})\right\},
\end{eqnarray}
\begin{eqnarray}\label{P_rho_z}
P^{(3)}_z(\mathbf{r},\rho_R)=6\chi^{(3)}_{xxyy}\left\{\frac{1}{1-\rho_R}\left[E^2_{p_z}(\mathbf{r})+\rho_R E^2_{p_x}(\mathbf{r})\right]E_{s_z}^*(\mathbf{r})+E_{p_x}(\mathbf{r})E_{p_z}(\mathbf{r})E_{s_x}^*(\mathbf{r})\right\}.
\end{eqnarray} 

 Given $\chi^{(3)}_{xxyy}$, $P^{(3)}_x(\mathbf{r})$ and $P^{(3)}_z(\mathbf{r})$ are now functions of $\mathbf{r}$, and $\rho_R$ only. Their dependence on $\mathbf{r}$ relies on the $E_{p_x}$, $E_{p_z}$, $E_{s_x}$ and $E_{s_z}$ field maps while $\rho_R$ only depends on the active medium.
 
 Eventually, throughout this paper, we assume no refractive index mismatch between the active medium and its environment (although it has been recently shown that refractive index mismatch can distort CARS radiation pattern \cite{Djaker06_ao}) while the active medium dispersion is assumed to be negligible (ie $n(\omega_p)=n(\omega_s)=n(\omega_{as})=1.33$).

\section{Mapping the components of the nonlinear polarisation}

As shown by Eqs.(\ref{P_rho_x}) and (\ref{P_rho_z}), the polarisations $P^{(3)}_x(\mathbf{r})$ and $P^{(3)}_z(\mathbf{r})$ tightly depend on (i) the spatial distribution of the fields components $E_{p_x}$, $E_{p_z}$, $E_{s_x}$ and $E_{s_z}$ and (ii) the depolarisation ratio $\rho_R$. We will be, first, interested in the behaviour of the exciting beams near the focal plane. Then, we will describe the induced nonlinear polarisation as a function of the focusing conditions and the depolarisation ratio $\rho_R$ of the active medium.\\

\subsection{Exciting fields}

Given the high numerical aperture of the objective, the exciting beams are diffraction-limited in the vicinity of the focal plane, following an Airy pattern. The spatial distribution of the fields components $E_{p_x}$, $E_{p_z}$, $E_{s_x}$ and $E_{s_z}$ only varies with the parameter $\beta$. As schematized on Figure \ref{fig:setup}, for any value of $\beta$, the depolarisation of the incident electric fields is maximal in the (\textit{xz})-plane and null in the (\textit{yz})-plane, so that both the pump and Stokes exciting fields along the \textit{z}-component are stronger in the former plane than in the latter. Consequently, for clarity, the study of the exciting fields (and a fortiori the induced nonlinear polarisation) will be restricted to the (\textit{xz})-plane (although it is computed everywhere).\\

\begin{figure}[!ht]
	\centering
		\includegraphics[width=1.0\textwidth]{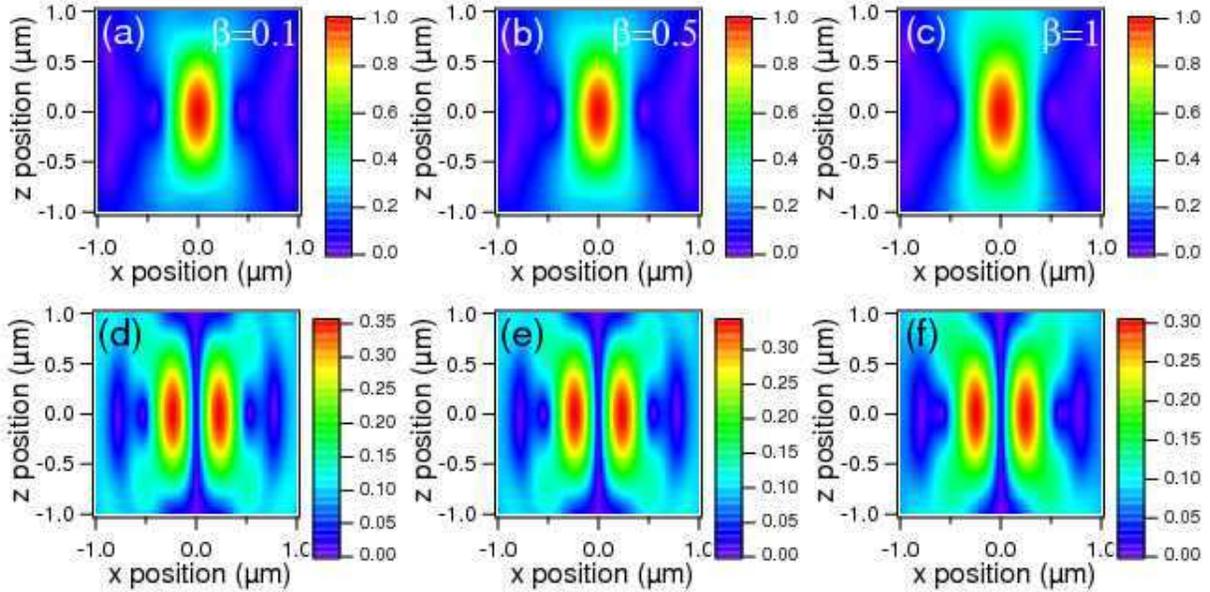}
	\caption{\small{normalised \textit{x-} (a-c)  and (d-f) \textit{z}-component amplitude maps of the pump exciting field in the vicinity of the focal plane. The paprameter $\beta$ equals 0.1 (a,d), 0.5 (b,e), and 1 (c,f). For each value of $\beta$, the amplitude of each component is normalised relative to the maximum of the total amplitude of the pump exciting field.}}
	\label{fig:champs_exc}
\end{figure}

Following Ref.\cite{Richards59}, the exciting fields are, excepted in some particular planes, elliptically polarised near the focus. However, throughout this part, we will only be interested in the amplitudes of the $E_{p_x}$, $E_{p_z}$, $E_{s_x}$ and $E_{s_z}$ components of the fields, as computed in Ref.\cite{Gachet06_spie}. The spatial distribution of the $E_{p_x}$ and $E_{p_z}$ components amplitude of the pump exciting field, near the focal plane, for different values of $\beta$ are depicted in Figure \ref{fig:champs_exc}. The $E_{p_z}$ component is rigorously null in the (\textit{yz})-plane, in agreement with Richards and Wolf \cite{Richards59}. Moreover, it is antisymmetric with respect to the focal point ($E_{p_z}(\mathbf{r})=E^*_{p_z}(-\mathbf r)$). When $\beta$ varies from 0.1 (a,d) to 1 (c,f), the tightness of the focusing decreases so that both the lateral and axial dimension of the focal volume increases (this effect is prevailing along the axial dimension). In parallel, the axial component $E_{p_z}$ gets lower. Such a behaviour can be easily explained by the ``filling'' of the objective back aperture by the incident pump beam. When $\beta$ equals 0.1, the incident pump beam overfills the microscope back aperture, so that it can be considered as a plane wave. On the contrary, when $\beta$ equals 1, it underfills the objective back aperture. $\beta$ equalling 0.5 (b,e) can be considered as realistic when the incident beam matches the objective back aperture, what is fulfilled in most experiments. The same conclusions can be drawn to the Stokes exciting field.\\

\subsection{Nonlinear polarisations}
 
 As shown previously, the parameter $\beta$ influences the exciting fields spatial distribution, and, therefore, the induced nonlinear polarisation. Furthermore, the observed medium itself influences the nonlinear polarisation via the Raman depolarisation ratio $\rho_R$.\\
 
 To study the influence of the observed active medium, we have considered three values for $\rho_R$ (0, 0.33 and 0.75). When $\rho_R$ equals 0, the Raman line is told to be totally polarised. Indeed, from Eqs.\ref{P_rho_x} and \ref{P_rho_z}, and the symbol $\cdot$ denoting the complex scalar product, the induced nonlinear polarisation is expressed by
 
\begin{eqnarray}\label{P_rho_0}
\mathbf{P}^{(3)}(\mathbf{r})=2\chi^{(3)}_{xxyy}\left(\mathbf{E_p}\cdot\mathbf{E_s}\right)\mathbf{E_p}
\end{eqnarray}

so that the induced nonlinear polarisation is oriented in the exciting pump field direction. $\rho_R$ equalling 0.33 corresponds to a midly polarised Raman line, where $\chi^{(3)}_{xxyy}=\chi^{(3)}_{xyyx}$. Eventually, in the case where $\rho_R$ equals 0.75, the Raman line is depolarised. Following Eqs.(\ref{P_rho_x}) and (\ref{P_rho_z}), a modification of the $\rho_R$ value affects the respective contributions of the $E_{p_x}$, $E_{p_z}$, $E_{s_x}$ and $E_{s_z}$ fields to the \textit{x}- and \textit{z}-components of the induced nonlinear polarisation. To make comparison between cases where $\rho_R$ varies, it is necessary to normalise the nonlinear polarisation distribution maps for each value of $\rho_R$.\\

To fully understand the normalisation procedure, it is important to note that, at the focus, the pump and Stokes exciting fields are only polarised along the \textit{x}-axis, ie $E_{p_z}$ and $E_{s_z}$ are rigorously null. Thus, following Eqs.(\ref{P_rho_x}) and (\ref{P_rho_z}), the induced nonlinear polarisation is oriented along the \textit{x}-axis too, so that at this point, it does not depend on $\rho_R$.\\

 Therefore, the expressions of the normalised components $P^{(3)}_{x\ norm}(\mathbf{r})$ and $P^{(3)}_{z \ norm}(\mathbf{r})$ are given by

\begin{eqnarray}\label{P_norm_2}
P^{(3)}_{x\ norm}(\mathbf{r})=\left|P^{(3)}_x(\mathbf{r})/P^{(3)}_x(\mathbf 0)\right|\ ,\ \ P^{(3)}_{z\ norm}(\mathbf{r})=\left|P^{(3)}_z(\mathbf{r})/P^{(3)}_x(\mathbf 0)\right|\ .
\end{eqnarray}

 A simple derivation of Eqs.(\ref{P_rho_x}) and (\ref{P_rho_z}) gives
 
\begin{eqnarray}\label{dP_rho_x}
\frac{\partial P^{(3)}_x(\mathbf{r})}{\partial \rho_R}=\frac{2\chi^{(3)}_{xxyy}}{1-\rho^2_R}\left[E_{p_x}^2(\mathbf{r})+E_{p_z}^2(\mathbf{r})\right] E_{s_x}^*(\mathbf{r}),
\end{eqnarray}
\begin{eqnarray}\label{dP_rho_z}
\frac{\partial P^{(3)}_z(\mathbf{r})}{\partial \rho_R}=\frac{2\chi^{(3)}_{xxyy}}{1-\rho^2_R}\left[E_{p_x}^2(\mathbf{r})+E_{p_z}^2(\mathbf{r})\right] E_{s_z}^*(\mathbf{r}).
\end{eqnarray}

 From Eqs.(\ref{dP_rho_x}) and (\ref{dP_rho_z}), it can be straightforward written

\begin{eqnarray}\label{dPz/dPx}
\left|\frac{\partial P^{(3)}_z(\mathbf{r})}{\partial \rho_R}/\frac{\partial P^{(3)}_x(\mathbf{r})}{\partial \rho_R}\right|=\left|\frac{E_{s_z}(\mathbf{r})}{E_{s_x}(\mathbf{r})}\right|.
\end{eqnarray}

\begin{figure}[!ht]
	\centering
		\includegraphics[width=1.\textwidth]{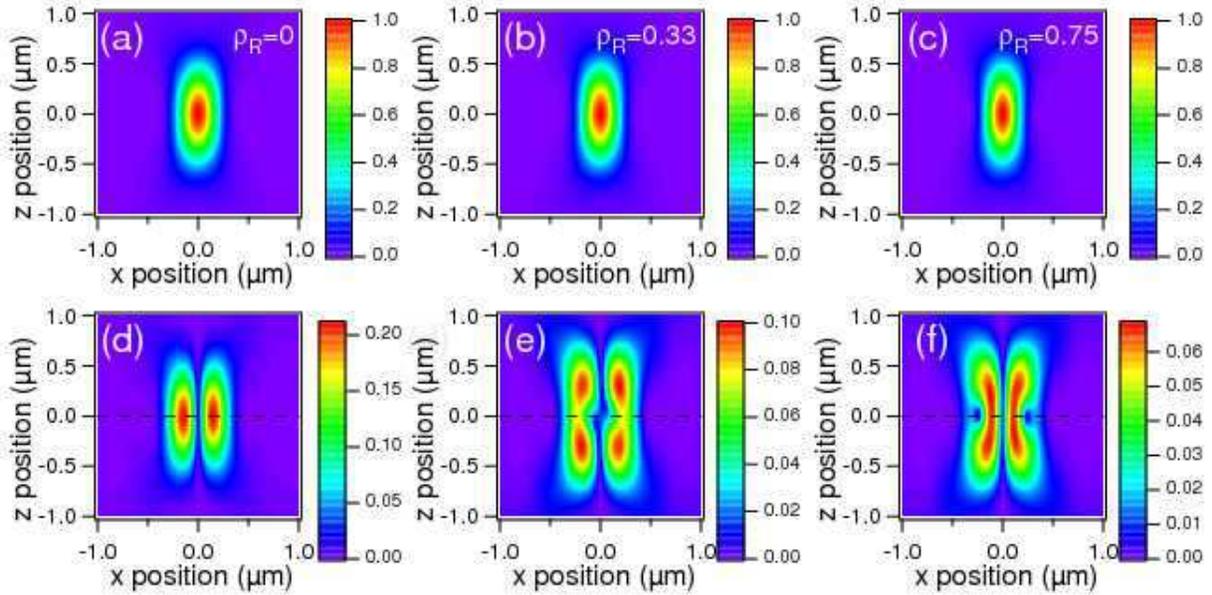}
	\caption{\small{normalised \textit{x-} (a-c)  and (d-f) \textit{z}-component amplitude maps of the nonlinear polarisation in the vicinity of the focal plane for $\beta=0.5$. The depolarisation ratio $\rho_R$ equals 0 (a,d), 0.33 (b,e), and 0.75 (c,f). For each value of $\rho_R$, the amplitude of each component is normalised relative to the amplitude of the nonlinear polarisation at the focus.}}
	\label{fig:Polar_0_5}
\end{figure}

In the vicinity of the focal plane, the $E_{s_x}$ amplitude being often stronger than the $E_{s_z}$ amplitude, the left side of Eq.(\ref{dPz/dPx}) often lies between 0 and 1. In other words, the amplitude of the $P^{(3)}_x$ component, at a given point $\mathbf{r}$, varies more quickly than the $P^{(3)}_z$ one with $\rho_R$. On Figure \ref{fig:Polar_0_5}, $P^{(3)}_{x\ norm}$ and $P^{(3)}_{z\ norm}$ are mapped in the (\textit{xz})-plane for $\beta=0.5$ and increasing values of $\rho_R$. While $P^{(3)}_{x\ norm}$ does not show any significant modification when $\rho_R$ varies from 0 (Figure \ref{fig:Polar_0_5}a) to 0.75 (Figure \ref{fig:Polar_0_5}c), the maximum of $P^{(3)}_{z\ norm}$ decreases from $0.22$ (Figure \ref{fig:Polar_0_5}d) to $0.07$ (Figure \ref{fig:Polar_0_5}f).\\

This decay is accompanied by a deformation of the  $P^{(3)}_{z\ norm}$ spatial distribution. Starting with two regular side lobes (Figure \ref{fig:Polar_0_5}d), it exhibits four lobes (Figure \ref{fig:Polar_0_5}e) and is finally butterfly-like-shaped (Figure \ref{fig:Polar_0_5}f).

\begin{figure}[!t]
	\centering
		\includegraphics[width=1.00\textwidth]{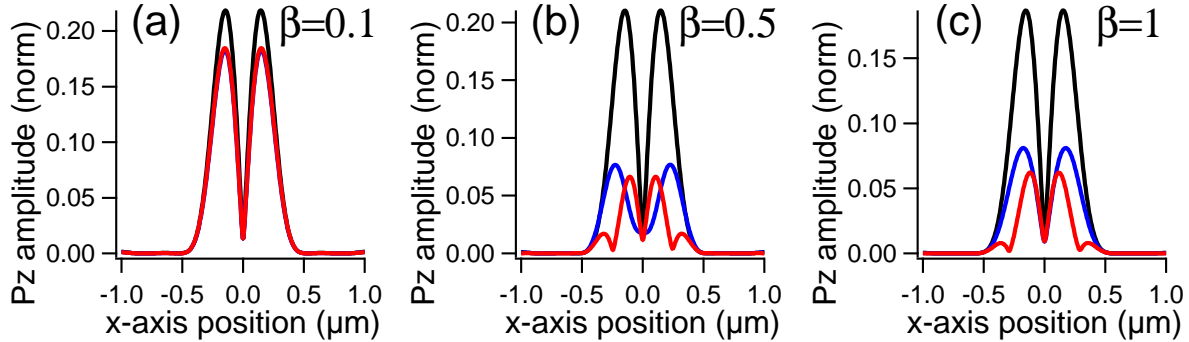}
	\caption{\small{normalised amplitude profiles of the nonlinear polarisation \textit{z}-component along the \textit{x}-axis when $\beta$ equals 0.1 (a), 0.5 (b) and 1 (c), for $\rho_R$ equaling 0 (dark traces), 0.33 (blue traces) and 0.75 (red traces). These profiles are taken along the dashed lines in Figure \ref{fig:Polar_0_5}.}}
	\label{fig:coupe_pz}
\end{figure}

To illustrate the influence of $\beta$ on the induced nonlinear polarisation, on Figure \ref{fig:coupe_pz} are dislayed $P^{(3)}_z$ normalised amplitude profiles along the \textit{x}-axis, for different values of $\beta$ and $\rho_R$ (the same profiles relative to the $P^{(3)}_x$ component are not plotted due to the weak influence of $\beta$ and $\rho_R$). This axis corresponds to the dashed lines on Figure \ref{fig:Polar_0_5}(d-f). It can be noted that in the case where incident plane waves are focused on the sample (smallest value of $\beta$), $\rho_R$ has only little influence on the induced nonlinear polarisation (Figure \ref{fig:coupe_pz}a). However, this influence gets stronger as $\beta$ grows (Figures \ref{fig:coupe_pz}b and \ref{fig:coupe_pz}c), although exciting fields intensity maps (see Figure \ref{fig:champs_exc}) exhibits only few modification.\\ 

\section{Consequences on CARS far-field radiation patterns}

The previous study has shown the dependency of the spatial evolution of the nonlinear induced polarisation to both the parameter $\beta$ and the Raman depolarisation ratio. Of course, as a coherent process, CARS generation is very sensitive to changes to the nonlinear polarisation. We show, in this part, how far-field radiation paterns are modified with $\rho_R$. As depicted in Figure \ref{fig:coupe_pz}, the nonlinear polarisation only slightly changes with $\rho_R$ when the incident exciting beams behave as plane waves ($\beta=0.1$). Therefore, the influence of the Raman depolarisation ratio to CARS radiation patterns has only been achieved for $\beta$ equalling 0.5 (the case where $\beta$ equals 1 seems to us quite far from usual experimental conditions).\\

Two kinds of objects have been investigated: thick and thin objects. The first class is predicted to only radiate in the forward (same direction of propagation as exicitng beams) direction while the second class also radiates in the epi direction (opposite direction of propagation to exciting beams) \cite{Cheng02_josa1}. Of course, fine features of these patterns depend on the exact shape of the objects. Following Figure \ref{fig:Polar_0_5}, the nonlinear polarisation is appreciable in the focal plane only in a $1µm \times 1µm$ wide square. For this reason, we find it relevant to assign this transverse dimension to both objects. They are taken as parallepipeds which axial extent (along the \textit{z}-axis) can be varied. The \textit{z}-component of the nonlinear polarisation being rigorously null at the exact focus, much smaller objects could not experience any modification of their radiation pattern with changing $\rho_R$. On the contrary, larger objects have their emitting area limited by the size of the excitation volume.\\

 To simplify the problem, we will first neglect the nonresonant part of the $\chi^{(3)}$ tensor. In a second part, we will take into account this contribution and show how it modifies the radiation patterns.\\
 
 \subsection{Purely resonant sample}

\begin{figure}[!ht]
	\centering
		\includegraphics[width=1.00\textwidth]{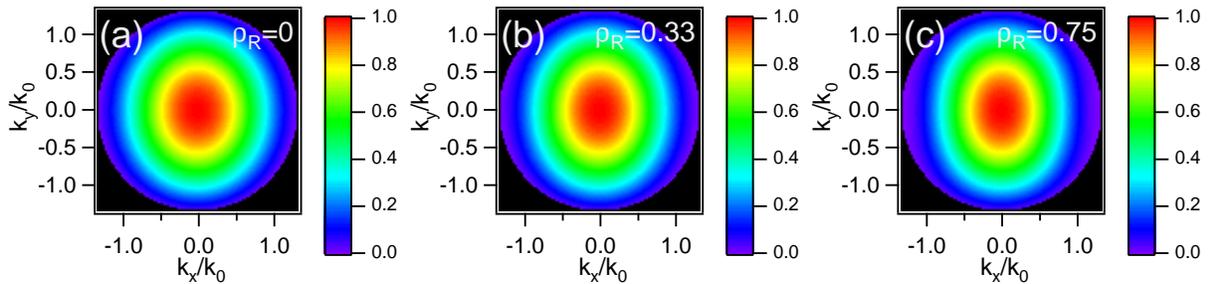}
	\caption{\small{Far-field radiation pattern in the reciprocal space ($k_x$,$k_y$) from a thick-shaped object ($1µm \times 1µm$, $500nm$ thick) centered in the (\textit{xy})-plane when $\rho_R$ equals 0 (a,d), 0.33 (b,e) and 0.75 (c,f), for $\beta=0.5$. Each diagram corresponds to forward-emitted signal. $k_0=2\pi/\lambda$. Each radiation pattern is normalised.}}
	\label{fig:bulk_k}
\end{figure}

Neglecting the nonresonant part of $\chi^{(3)}$, we start with a $500nm$ thick object. Its forward radiation patterns, in the reciprocal space ($k_x$,$k_y$), are displayed on Figure \ref{fig:bulk_k}, for $\rho_R$ lying between 0 (a) and 0.75 (c). In this case, the Raman depolarisation ratio has little effect on the far-field CARS radiation pattern. The only feature to be noted is the slightly decreasing divergence of the anti-Stokes beam in the (\textit{yz})-plane as $\rho_R$ increases.\\

\begin{figure}[!hb]
	\centering
		\includegraphics[width=1.00\textwidth]{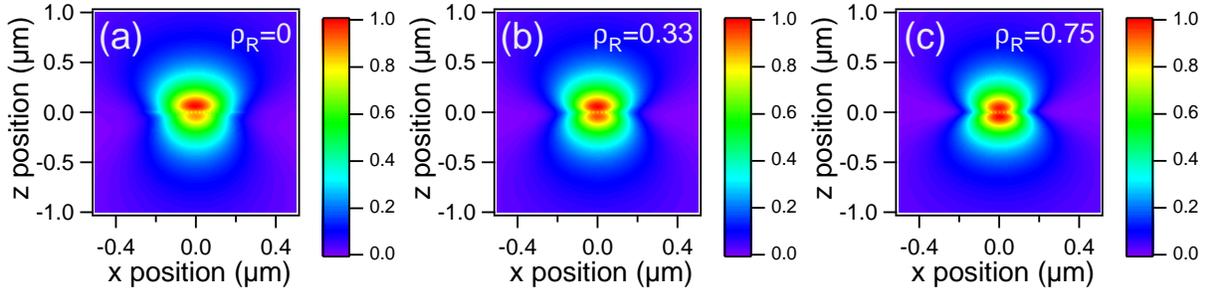}
		\caption{\small{Far-field radiation pattern in the (\textit{xz})-plane from a slice-shaped object ($1µm \times 1µm$) located in the (\textit{xy})-plane when $\rho_R$ equals 0 (a), 0.33 (b) and 0.75 (c), for $\beta=0.5$. Each radiation pattern is normalised.}}
	\label{fig:nappe_direct}
\end{figure}

\begin{figure}[!ht]
	\centering
		\includegraphics[width=1.00\textwidth]{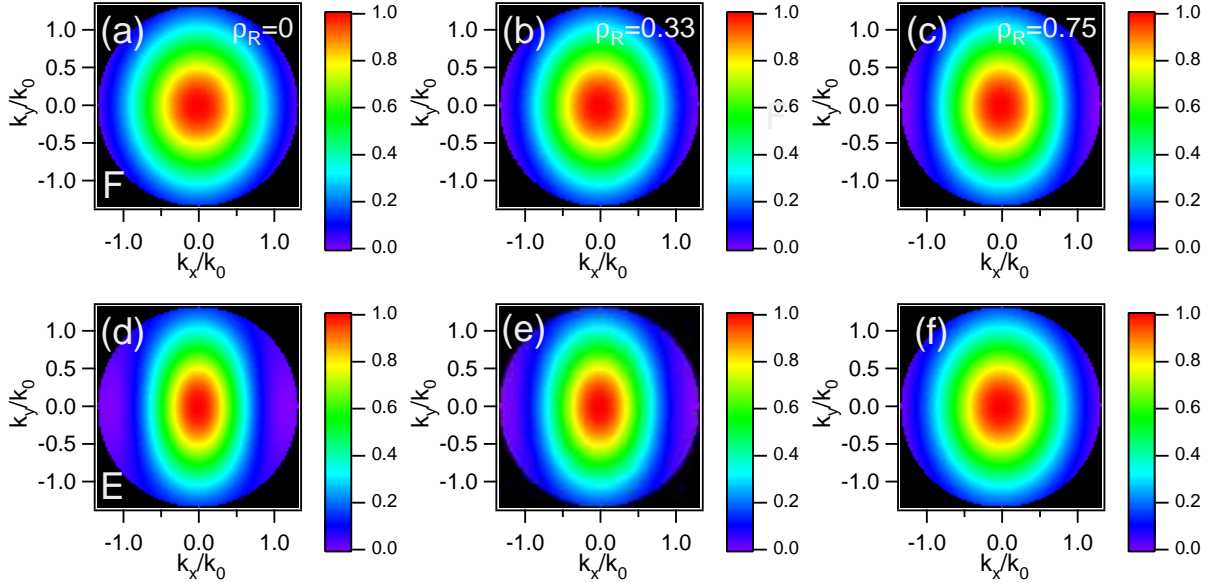}
		\caption{\small{Far-field radiation pattern in the reciprocal space ($k_x$,$k_y$) from a slice-shaped object ($1µm \times 1µm$) located in the (\textit{xy})-plane when $\rho_R$ equals 0 (a,d), 0.33 (b,e) and 0.75 (c,f), for $\beta=0.5$. (a-c) forward-emitted signal (F); (d-f) epi-emitted signal (E). $k_0=2\pi/\lambda$. For each value of $\rho_R$, each radiation pattern is normalised relative to the forward intensity at ($k_x=0$,$k_y=0$).}}
	\label{fig:nappe_k}
\end{figure}

\begin{figure}[!ht]
	\centering
		\includegraphics[width=0.70\textwidth]{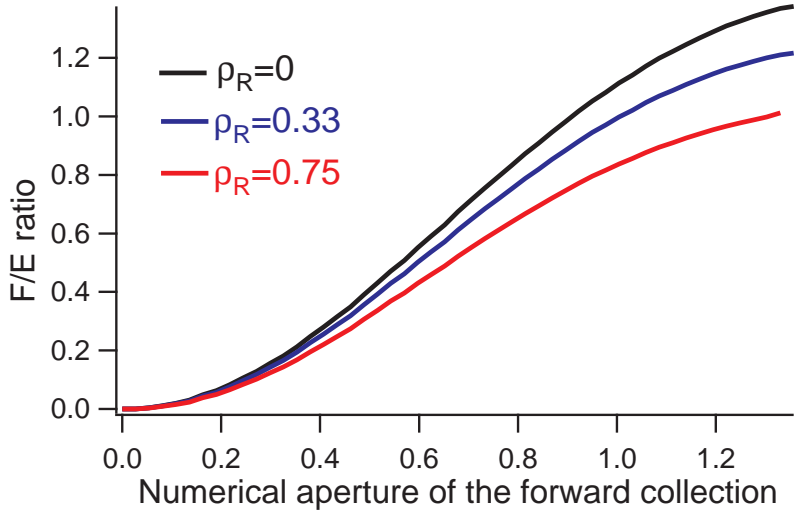}
		\caption{\small{Forward on epi detected intensity ratio (F/E ratio) as a function of the numerical aperture of the forward collection for $\rho_R$ equalling 0 (black), 0.33 (red) and 0.75 (blue). For each value of $\rho_R$, the ratio is normalised with respect to the epi-emitted signal intensity collected with a 1.2 NA objective.}}
	\label{fig:collection}
\end{figure}

The forward and epi radiation patterns of a slice-shaped object, for various values of $\rho_R$, in the (\textit{xz})-plane and in the reciprocal space ($k_x$,$k_y$) are shown on Figures \ref{fig:nappe_direct} and \ref{fig:nappe_k}. The object is now an infinitely thin slice of dipoles. It is morphologically identical to biological membranes found in cells. Following Figure \ref{fig:nappe_direct}, the radiation pattern tends to be symmmetrical when $\rho_R$ approaches 0.75 (Figure \ref{fig:nappe_direct}c). Reminding previous results obtained for the induced nonlinear polarisation (Figure \ref{fig:Polar_0_5}), the observed symmetry conveys the \textit{x}-orientation of the dipoles. Further information is drawn from Figure \ref{fig:nappe_k}. First, as in the case of the thick medium, a very slight change in the forwardly-emitted anti-Stokes beam divergence is observed (Figure \ref{fig:nappe_k}a-c). Then, a more important change in the divergence of the epi-emitted anti-Stokes beam occurs in the (\textit{xy})-plane (Figure \ref{fig:nappe_k}d-f). Contrarly to the case of the forwardly-emitted beam, the divergence increases with $\rho_R$. Figure \ref{fig:collection} displays the ratio of forward to epi-collected intensity as a function of the forward-collection NA (noted F/E ratio). The epi-collection NA is supposed to be constant and to equal the excitation NA, ie 1.2, and the forward-collected intensity is normalised relative to the epi-collected intensity. Naturally, for any value of $\rho_R$, the F/E ratio is an increasing function of the forward-collection NA. Moreover, for any value of the forward-collection NA, the higher the $\rho_R$ value, the smaller the F/E ratio. A further analysis shows a relative variation of the F/E ratio lying between $26\%$ and $36\%$ when $\rho_R$ varies from 0 to 0.75. It lies around $26\%$ for low NA (typically less than 0.3). For commonly used 0.5 NA condensors, it equals $28\%$ and when the collection is insured by another high NA objective (1.2 in water for example), it reaches $35\%$.\\

\subsection{Influence of the nonresonant part}\label{NR_influence}

The previous investigation presented the advantage to physically highlight the main modifications of the raditaion pattern with the parameter $\rho_R$. However, it failed to depict a realistic CARS experiment as it neglected the nonresonant part $\chi^{(3)}_{NR}$ of the nonlinear tensor. To take it into account, we have considered the previous ``thin'' object, located in the (\textit{xy})-plane, whose third order nonlinear polarisation is now given by

\begin{eqnarray}\label{NR_influence_eqn}
\mathbf{P}^{(3)}(\mathbf{r})=a\cdot\mathbf{P}^{(3)}(\mathbf{r},\rho_R=1/3)+(1-a)\cdot\mathbf{P}^{(3)}(\mathbf{r},\rho_R=0)\cdot \exp(i\,\pi/2)
\end{eqnarray}

where $\mathbf{P}^{(3)}(\mathbf{r},\rho_R)$ is defined by its \textit{x}- and \textit{z}-components in Eqs.\ref{P_rho_x} and \ref{P_rho_z} and \textit{a} is a weighting coefficient.\\

$\mathbf{P}^{(3)}(\mathbf{r},\rho_R=1/3)$ stands for the nonresonant contribution while $\mathbf{P}^{(3)}(\mathbf{r},\rho_R=0)$ stands for the resonant contribution (note the $\pi/2$ dephasing, with respect to the nonresonant contribution, at resonance). $\rho_R=0$ was chosen for the resonant contribution since it exhibits the strongest F/E asymmetry.\\

\begin{figure}[!ht]
	\centering
		\includegraphics[width=0.70\textwidth]{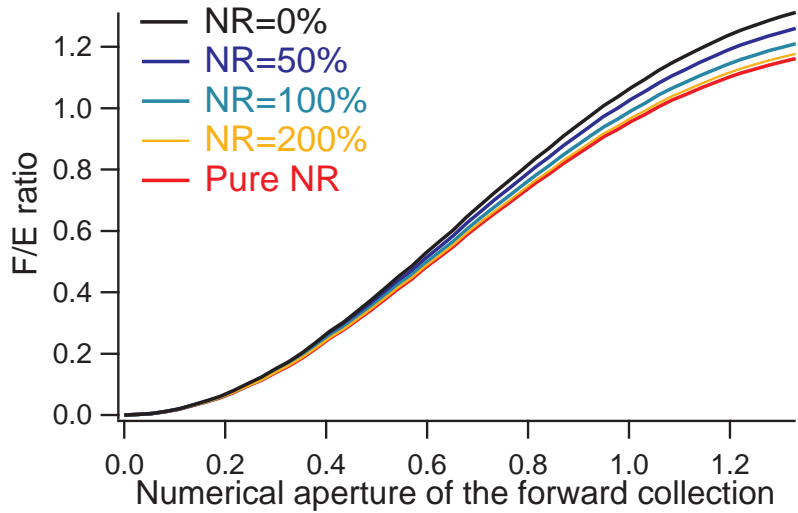}
		\caption{\small{F/E intensity ratio as a function of the numerical aperture of the forward collection for $\rho_R=0$. The relative nonresonant part of the nonlinear tensor varies from 0 (black) to 2 (orange). In red, the case of a thin object with $\rho_R$ which value equals $1/3$. For each value of the nonresonant part NR, the ratio is normalised with respect to the epi-emitted signal intensity collected with a 1.2 NA objective.}}
	\label{fig:collection2}
\end{figure}

 Starting with only the resonant part ($\textit{a}=0$), the nonresonant part was increased from a tenth of the resonant part ($\textit{a}=1/11$) to twice ($\textit{a}=2/3$). The intensity ratii F/E, for these values of the nonresonant part (NR), as a function of the numerical aperture of the forward collection are plotted on Figure \ref{fig:collection2}. As expected, the epi and forward radiation patterns of this object come closer to those of a ``purely nonresonant object'' (ie which Raman depolarisation ratio equals $1/3)$ with increasing contribution of the nonresonant part. The nonresonant part thus attenuates the slight differences observed in radiation patterns for changing values of the Raman depolarisation ratio. This is also true for objects with various shapes and Raman depolarisation ratii.\\

 In biological samples, where the imaged samples are surrounded by solvent such as water, we predict (for thin objects) that the forward radiation pattern is governed by the Raman depolarisation $1/3$ of the nonresonant surrounding medium while the epi radiation pattern is driven by both the Raman depolarisation ratio of the object and the relative strength of its nonresonant part.\\

\section{Conclusion}

Through this paper, a further investigation of far-field CARS radiation patterns under tight focusing conditions has been lead through a full-vectorial study. It has revealed the conjoined role of focusing conditions (through the parameter $\beta$) and the Raman depolarisation ratio $\rho_R$ of the studied medium, in addition of those, already known, of the size and shape of imaged objects. While the far-field radiation pattern of thick object is not affected by changes in value of $\beta$ and $\rho_R$, those of thin objects are slightly modified, concerning both the anti-Stokes beam divergence and the ratio of epi to forward-generated power. Such effects cannot be seen when neglecting the longitudinal components of the exciting fields. However, they might be only observable for strong Raman lines (such as the relative nonresonant part is weak) which is not always the case, specially when working with biological samples. In most cases, this study validates the treatment of the problem previously proposed by Cheng and al. \cite{Cheng02_josa1}. However, in the case of thin objects, it brings some corrections. Such objects are encountered when imaging biological samples, cellular membranes being a few nanometers thick.\\

 We have restricted the analysis to the case of isotropic media, excited with collinearly polarised exciting beams. Furthermore, the non-resonant surrounding solvent or matrix has been neglected in the computations but its effect can be easily predicted from the nonresonant contribution analysis (see section \ref{NR_influence}). The situation is far more complex when taking into account the anisotropy of studied media as well as the possible ellipticity of the exciting beams polarisations. It can be, of course, modelled, following the same electromagnetic treatment. For the case of electronically resonant CARS \cite{Druet78_pr}, the Raman depolarisation ratio no longer lies between 0 and 0.75 and varies on a larger range \cite{Otto01_jrs}, so that the situation must be reexamined very carefully.\\
 
\section*{Acknowledgements}

One of us (DG) acknowledges a grant from the French Ministry for National Education, Research and Technology. This work is supported by the Centre National de la Recherche Scientifique (CNRS) and the European Union (through the FEDER program).


\end{document}